\documentclass[%
reprint,
superscriptaddress,
nofootinbib,
amsmath,amssymb,
pra,
]{revtex4-1}

\pdfoutput=1
\usepackage{float}
\usepackage{xcolor}
\usepackage{graphicx}
\usepackage{dcolumn}
\usepackage{bm,lipsum}
\usepackage[colorlinks=true, linkcolor=blue, citecolor=blue, urlcolor=blue]{hyperref}
\newcommand{\ZIB}{Zuse Institute Berlin, 14195 Berlin, Germany}
\newcommand{\JCM}{JCMwave GmbH, 14050 Berlin, Germany}

\usepackage{graphicx}
\usepackage{booktabs}
\usepackage{tabularx}
\usepackage{xcolor,colortbl}

\usepackage{pdfpages}
\makeatletter
\AtBeginDocument{\let\LS@rot\@undefined}
\makeatother

\newcommand{\mytoprule}{\specialrule{0.1em}{0.2em}{0.2em}}
\newcommand{\mymidrule}{\specialrule{0.1em}{0.2em}{0.2em}}
\newcommand{\mybottomrule}{\specialrule{0.1em}{0.2em}{0.2em}}

\definecolor{Gray}{gray}{0.9}

\newcolumntype{a}{>{\columncolor{Gray}}c}
\newcolumntype{b}{>{\columncolor{white}}c}

\begin{document}

\title{Poles and zeros in non-Hermitian systems: Application to photonics}

\author{Felix Binkowski}
\affiliation{\ZIB}
\author{Fridtjof Betz}
\affiliation{\ZIB}
\author{Rémi Colom}
\affiliation{Université Côte d’Azur, CNRS, CRHEA, 06560 Valbonne, France}
\author{Patrice Genevet}
\affiliation{Université Côte d’Azur, CNRS, CRHEA, 06560 Valbonne, France}
\affiliation{Physics Department, Colorado School of Mines, Golden, Colorado 80401, USA} 
\author{Sven~Burger}
\affiliation{\ZIB}
\affiliation{\JCM}

\begin{abstract}
Resonances are essential for understanding the interactions between light
and matter in photonic systems. The real frequency response of the non-Hermitian systems depends on the
complex-valued resonance frequencies, which are the poles of electromagnetic response functions.
The zeros of the response functions are often used for designing devices,
since the zeros can be located close to the real axis, where they have
significant impact on scattering properties. While methods are available to
determine the locations of the poles, there is a lack of appropriate
approaches to find the zeros in photonic systems.
We present an approach to compute poles and zeros based on contour
integration of electromagnetic quantities. This also allows to extract
sensitivities with respect to geometrical or other parameters enabling
efficient device design. The approach is
applied to a topical example in nanophotonics, an illuminated metasurface,
where the emergence of reflection zeros due to the underlying resonance poles
is explored using residue-based modal expansions. The generality and simplicity of the theory
allows straightforward transfer to other areas of physics. We expect that
easy access to zeros will enable new computer-aided design
methods in photonics and other fields.
\end{abstract}

\maketitle
\section{Introduction}
In the field of photonics, light-matter interactions can be tuned by exploiting resonance phenomena.
Examples include tailoring quantum entanglement with atoms and photons in cavities~\cite{Haroche_RevModPhys_2001},
probing single molecules with ultrahigh sensitivity~\cite{Nie_Science_1997}, and
realizing efficient single-photon sources~\cite{Senellart_2017}.
While electromagnetic observables are measured at real-valued excitation frequencies, 
the concept of resonances intrinsically considers
the complex frequency plane~\cite{Lalanne_QNMReview_2018,Zworski_Scattering_Resonances_2019}.
Resonance frequencies are complex-valued as the systems are non-Hermitian, e.g., 
due to interaction with the environment~\cite{Wu_ACSPhot_2021}. 
Resonances are natural properties of photonic systems, often featuring highly localized
electromagnetic field intensities, and correspond to poles of electromagnetic response functions,
such as scattering ($S$) matrix, reflection, or transmission coefficients.
Resonances can also serve as a basis for the expansion of the response functions.
Although most nanophotonic systems support many resonances, often only a few
resonances are sufficient to determine the optical 
response in the real-valued frequency range of interest~\cite{Sauvan_2022,Nicolet_2022}.
\footnote{This work has been published:\\
F. Binkowski et al., Phys. Rev. B \textbf{109}, 045414 (2024).\\
DOI: \href{https://doi.org/10.1103/PhysRevB.109.045414}{10.1103/PhysRevB.109.045414}}

\begin{figure*}[]
\includegraphics[width=0.98\textwidth]{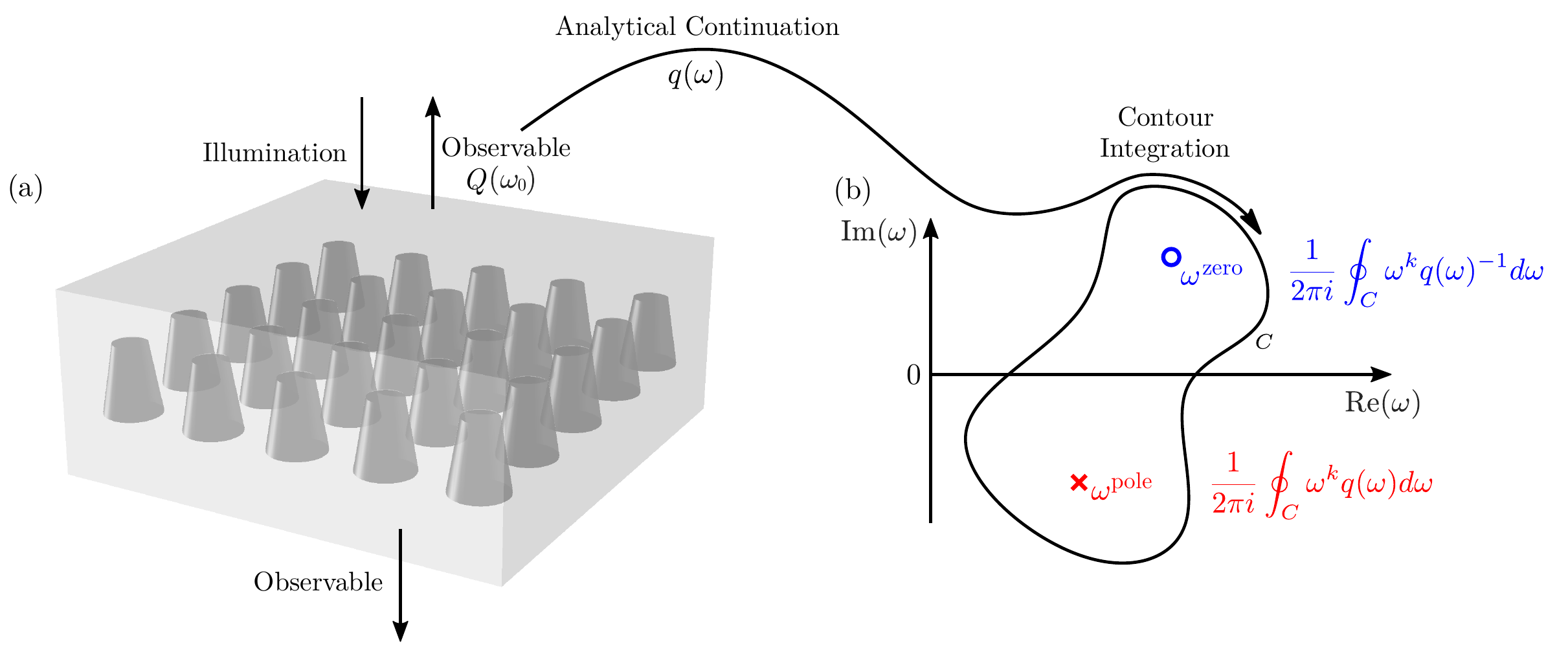}
\caption{\label{fig01} 
Singularities of physical observables emerge in the complex frequency plane. 
(a) Schematic of a metasurface illuminated by a plane wave.
The response of the nanostructure is typically described by resulting physical observables $Q(\omega_0)$
for real excitation frequencies $\omega_0 \in \mathbb{R}$.
The corresponding analytical continuation $q(\omega)$ into the complex 
frequency plane $\omega \in \mathbb{C}$ is considered.
(b) Both quantities $q(\omega)$ and $q(\omega)^{-1}$ exhibit 
singularities in the complex frequency plane. 
A singularity of $q(\omega)$ is denoted by $\omega^\mathrm{pole}$ 
and a singularity of $q(\omega)^{-1}$ is denoted by $\omega^\mathrm{zero}$.
There exists a relation between contour integrals involving $q(\omega)$ and $q(\omega)^{-1}$, respectively,
and their singularities. This allows for determining the locations
of $\omega^\mathrm{pole}$ and~$\omega^\mathrm{zero}$.}
\end{figure*}

Photonic response functions not only have poles but also zeros
describing the vanishing of the response functions for the considered input-output channels.
Just like the poles in non-Hermitian systems, the zeros generally occur in the complex frequency plane.
For example, in systems without absorption, the zeros of the $S$-matrix coefficients
are complex conjugates of the underlying poles~\cite{Nussenzveig_1972}. 
In the case of reflection or transmission coefficients,
poles and zeros do not necessarily occur as complex conjugated pairs.
In particular, in the absence of absorption, 
it is possible that zeros lie exactly on the real axis~\cite{Sweeney_2020,Kang_2021,Colom_2023}.
Zeros are equally important for the qualitative prediction of the real frequency response,
even if they occur at complex-valued frequencies.
Therefore, controlling the relative locations of the poles and zeros in the complex
frequency plane can be considered as an alternative, more fundamental approach to design
photonic systems in general.

This kind of approach has long been used to design electronic systems~\cite{Desoer_1974}.
For example, all-pass filters, i.e., systems whose response amplitude remains
constant when the excitation frequency is varied, have poles
and zeros which are complex conjugates of each other~\cite{Oppenheim_2017}.
Other examples are minimum-phase systems, where the zeros have to be restricted
to the lower part of the complex plane~\cite{Bechhoefer_2011}.
In photonic crystals, bound states in the continuum can exist when a
pole and a zero of the $S$-matrix coincide on the real axis~\cite{Hsu_BICs_NatRevMat_2016}.
Away from this condition, the pole and zero split and may occur in the 
complex frequency plane~\cite{Sakotic_2023}. Exceptional points,
which have recently attracted much attention in photonics
due to their potential for sensing~\cite{Wiersig_2020},
occur when at least two poles or two zeros merge 
yielding a higher order singularity~\cite{Miri_2019,Sweeney_2019,Moritake_2023}.
It has further been shown that a $2\pi$-phase gradient of the reflection or
transmission output channel of a metasurface can be realized when a pair of pole and zero
is separated by the real axis~\cite{Colom_2023} and
that phase gradient metasurfaces can be designed by exploiting the $2\pi$-phase winding
around zeros of cross-polarization reflection coefficients \cite{Song_2021}.
The zeros of photonic systems can have arbitrarily small imaginary parts,
i.e., the analysis of the locations of the zeros is extremely relevant
to design the response of the systems at real frequencies. 
Total absorption of light or perfect coherent absorption occurs when zeros of the
$S$-matrix are on the real axis~\cite{Hutley_1976,Chong_2010,Maystre_2013}.
Reflection zeros are also exploited for phase-sensitive detection with
nanophotonic cavities in biosensing applications~\cite{Sreekanth_2018,Kravets_2018}.

While in many electronic systems the determination of poles and zeros of the transfer matrix
may be done analytically, this is often not possible for photonic structures.
To compute zeros of $S$-matrix, reflection, and transmission coefficients of specific systems, 
it has been proposed to solve Maxwell's equations as an eigenproblem with
appropriately modified boundary conditions~\cite{Shao_1995, Grigoriev_2013,Dhia_2018, Sweeney_2020}.

In this work, we develop a framework for the investigation of poles and zeros of electromagnetic
response functions, such as $S$-matrix, reflection, or transmission coefficients, in non-Hermitian systems.
The underlying approach exploits a contour integral method well-known in numerical mathematics.
The framework, based on scalar physical quantities,
extends this approach and enables the simultaneous determination of poles, zeros, sensitivities,
and residue-based modal expansions.
This is demonstrated by a numerical investigation of the reflection of a photonic metasurface.
The occurrence of reflection zeros due to the interference of modal contributions corresponding to poles is observed.

\section{Singularities and contour integration} In the steady-state regime,
light scattering in a material system can be described 
by the time-harmonic Maxwell's equation in second-order form,
\begin{align}
\begin{split}
	\nabla \times \mu^{-1} 
	&\nabla \times \mathbf{E} -
	\omega_0^2\epsilon \mathbf{E}  = 
	i\omega_0\mathbf{J}, \label{eq:Maxwell}
 \end{split}
\end{align}
where $\mathbf{E}(\mathbf{r},\omega_0) \in \mathbb{C}^3$ is the electric field,
$\mathbf{J}(\mathbf{r})\in \mathbb{C}^3$ is 
the electric current density describing a light source,
$\epsilon(\mathbf{r},\omega_0)$ and $\mu(\mathbf{r},\omega_0)$ are the complex-valued
permittivity and permeability tensors, respectively,
$\mathbf{r}\in\mathbb{R}^3$ is the position, and $\omega_0\in\mathbb{R}$ is the angular frequency.

\begin{figure*}[]
\includegraphics[width=0.98\textwidth]{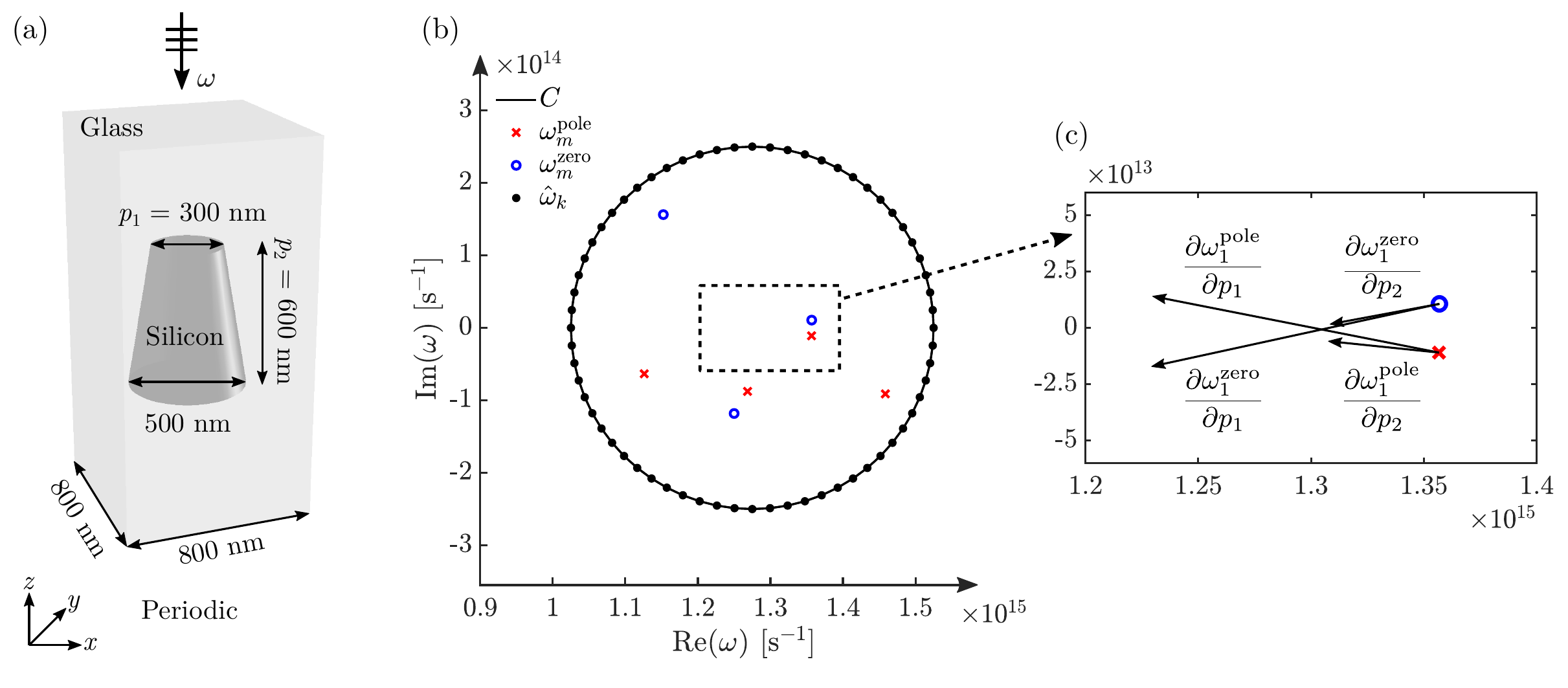}
\caption{\label{fig02}
Poles, reflection zeros, and their sensitivities of an illuminated metasurface.
(a) Sketch of the unit cell, which is periodic in $x$ and $y$ direction.
The metasurface consists of silicon cones with 
upper radius $p_1$ and height $p_2$ embedded in an infinitely extended glass medium.
The refractive indices of silicon and glass are $n = 3.5$ and  $n = 1.5$, respectively.
The illumination is given by a $x$ polarized plane wave of
the optical frequency $\omega$ at normal incidence from above. 
(b) Integration contour $C$, which is a circle with center 
$1.275 \times 10^{15} \, \mathrm{s}^{-1}$ and radius
$2.5 \times 10^{14} \, \mathrm{s}^{-1}$.
The integration contour is numerically discretized by a trapezoidal rule with
$64$ integration points $\hat{\omega}_k$ marked with black dots.
The resulting poles $\omega^{\mathrm{pole}}_m$ and reflection zeros $\omega^{\mathrm{zero}}_m$
correspond to the Fourier coefficient $q(\omega)$. 
(c) Detail of the complex frequency plane comprising the two singularities
$\omega^{\mathrm{pole}}_1$ and $\omega^{\mathrm{zero}}_1$.
The arrows are associated to the corresponding sensitivities
$\partial \omega^{\mathrm{pole}}_1 / \partial p_k$ and
$\partial \omega^{\mathrm{zero}}_1 / \partial p_k$
with respect to the shape parameters $p_1$ and $p_2$.
The computation of the sensitivities is also based on integrals along the contour $C$.
The complex-valued arrows are given by 
$\delta \omega = \partial \omega / \partial p \times 100 \mathrm{nm}$.}
\end{figure*} 

Electromagnetic quantities $Q(\mathbf{E}(\mathbf{r},\omega_0)) \in \mathbb{C}$
are typically experimentally measured 
for real excitation frequencies \mbox{$\omega_0 \in \mathbb{R}$}.
However, 
to obtain deeper insights into light-matter interactions in non-Hermitian systems,
an investigation of the optical response for complex
frequencies $\omega \in \mathbb{C}$ is essential.
For this, we consider the analytical continuation of
$Q(\mathbf{E}(\mathbf{r},\omega_0))$ into the complex frequency plane,
which we denote by $q(\omega) \in \mathbb{C}$ as a short notation of $q(\mathbf{E}(\mathbf{r},\omega))$.
Figure~\ref{fig01}(a) shows an example from the field of nanophotonics,
a dielectric metasurface~\cite{Genevet_2023}. Illumination of the metasurface by a plane wave
with the optical frequency $\omega_0$ yields a physical observable $Q(\omega_0)$.
The singularities of its analytical continuation $q(\omega)$
and the singularities of $q(\omega)^{-1}$ are of special interest and
can be used to investigate the properties of the metasurface.

The singularities of $q(\omega)$ are the
poles $\omega^\mathrm{pole}$ of the physical quantity $q(\omega)$.
The associated electric fields are so-called resonances or quasinormal modes,
which are also solutions of Eq.~\eqref{eq:Maxwell} without a source term
and with losses, e.g., due to open boundaries or dissipation in the system.
The singularities of $q(\omega)^{-1}$ are the zeros $\omega^\mathrm{zero}$ of $q(\omega)$.
The associated electric fields lead to $q(\omega^\mathrm{zero}) = 0$.
Figure~\ref{fig01}(b) sketches the complex frequency plane with exemplary locations of a pole and a zero.
By using Cauchy's integral theorem for a contour $C$ which
encloses one simple pole $\omega^{\mathrm{pole}}$
and (or) one simple zero $\omega^{\mathrm{zero}}$ of the quantity $q(\omega)$, as sketched in Fig.~\ref{fig01}(b),
$\omega^{\mathrm{pole}}$ and $\omega^{\mathrm{zero}}$ are given by
\begin{align}
  \omega^{\mathrm{pole}} = \frac{\oint \nolimits_{{C}}\omega q(\omega)
    d\omega}{\oint \nolimits_{{C}}q(\omega)
    d\omega} \hspace{0.25cm} \mathrm{and} \hspace{0.25cm}
   \omega^{\mathrm{zero}} = \frac{\oint \nolimits_{{C}}\omega q(\omega)^{-1}
    d\omega}{\oint \nolimits_{{C}}q(\omega)^{-1}
    d\omega}, \nonumber
\end{align}
respectively.
The locations of $M$ poles $\omega^\mathrm{pole}_{m}$
inside a contour $C$ are given by the eigenvalues $\omega_m$ of the
generalized eigenproblem~\cite{Austin_2014}
\begin{align}
H^{<} X = H X \Omega, \label{eq:Hankel}
\end{align}
where $\Omega = \mathrm{diag}(\omega_1,\dots,\omega_M)$
is a diagonal matrix containing the eigenvalues, the columns of the
matrix $X \in \mathbb{C}^{M\times M}$ are the eigenvectors, and
\begin{align}
	H =
	\begin{bmatrix}
	s_0	& \dots & s_{M-1} 		\\
	\vdots 	&   		 & \vdots	\\
	s_{M-1}		&\dots 	& s_{2M-2}		
	\end{bmatrix}, \,\,\,
  	H^{<} =
	\begin{bmatrix}
	s_1	& \dots & s_{M} 		\\
	\vdots 	&    	 & \vdots	\\
	s_{M}		&\dots 	& s_{2M-1}		
        \end{bmatrix} \nonumber
\end{align}
are Hankel matrices with the contour-integral-based elements
\begin{align}
s_k = \frac{1}{2\pi i} \oint \nolimits_{{C}} \omega^k q(\omega)
     d\omega.  \nonumber  
\end{align}
The zeros $\omega^{\mathrm{zero}}_{m}$ inside the contour are also given in this way, except that
the quantity $q(\omega)^{-1}$ is considered for the elements instead of $q(\omega)$.
Note that this type of approach has inspired a family of numerical methods
to reliably evaluate all zeros and poles in a given bounded domain.
The methods are an active area of research in numerical mathematics, where, e.g., numerical stability,
error bounds, and adaptive subdivision schemes are 
investigated~\cite{Delves_1967,Kravanja_2000,Austin_2014,Chen_2022}.
In the field of photonics, poles are usually determined by computing
the quasinormal modes as electromagnetic vector fields, 
where, e.g., the Arnoldi~\cite{Saad_Book_NumMeth_Eig_2011,Yan_PRB_2018},
FEAST~\cite{Gavin_JCompPhy_2018}, or Beyn's~\cite{Beyn_LAAppl_2012} algorithm is used.
This is in contrast to the approach presented in this work,
where scalar physical quantities are considered.

To compute poles and zeros, the elements of the Hankel
matrices can be approximated by numerical integration~\cite{Trefethen_SIAM_Trapz_2014},
where the quantity of interest $q(\omega)$ is calculated
by computing $\mathbf{E}(\mathbf{r},\omega)$ for complex frequencies
on the integration contour $C$.
The electric field $\mathbf{E}(\mathbf{r},\omega)$ can be obtained by numerically 
solving Maxwell's equation given in Eq.~\eqref{eq:Maxwell}.
In general, the electric field is not meromorphic everywhere in the complex plane,
due to branch cuts and accumulation points~\cite{Sauvan_2022}. For our approach,
the electric field must be analytic only in the spectral region of interest, except at the poles to be investigated.
The quantity $q(\omega)^{-1}$ is immediately available by inverting the scalar quantity $q(\omega)$.
Computing the different contour integrals for each of the elements requires no additional computational
effort since the quantity $q(\omega)$ needs to
be calculated only once for each of the integration points.
The integrands differ only in the weight functions $\omega^k$.
Information on the numerical realization can be found 
in Sec.~S1 in the Supplemental Material~\cite{Supplement_Poles_Zeros} and in Ref.~\cite{Betz_2021}.
Further, the data publication~\cite{Binkowski_SourceCode_Poles_Zeros} contains software 
for reproducing the results of this work, based on an interface to the finite-element-based
Maxwell solver JCMsuite.  

\renewcommand{\arraystretch}{1.6}
\begin{table}[]
	\caption{Computed pole $\omega^\mathrm{pole}_1$ and zero $\omega^\mathrm{zero}_1$ in $\mathrm{s}^{-1}$ and
            their sensitivities $\partial \omega^\mathrm{pole}_1 / \partial p_k$ and
                $\partial \omega^\mathrm{zero}_1 / \partial p_k$ in $(\mathrm{s\,nm})^{-1}$.
                The sensitivities are computed at the parameter reference values shown in  Fig.~\ref{fig02}(a)
                using the $64$ integration points on the contour $C$ depicted in Fig.~\ref{fig02}(b).
                The relative error is defined by
                $\mathrm{err}(\mathrm{Re}(u)) = |(\mathrm{Re}(u) - \mathrm{Re}(u_\mathrm{ref})) / \mathrm{Re}(u_\mathrm{ref})|$,
                where $u_\mathrm{ref}$ is the reference solution computed with $256$ integration points.
                }
	\begin{tabularx}{0.495\textwidth}{c  a  b  a  b} \mytoprule
		  $u$ 
		& $\mathrm{Re}(u)$
		& $\mathrm{Im}(u)$
            & \hspace{0.075cm} $\mathrm{err}(\mathrm{Re}(u))$ \hspace{0.075cm}
            & \hspace{0.075cm} $\mathrm{err}(\mathrm{Im}(u))$  \\
		\mymidrule
            $\omega^\mathrm{pole}_1$ & $1.357\times10^{15}$ & $-1.095\times 10^{13}$
            & $ < 10^{-8}$ & $3\times 10^{-8}$ \\ 
            $\frac{\partial \omega^\mathrm{pole}_1}{\partial p_1}$ & $-1.257\times 10^{12}$ & $2.47\times 10^{11}$ 
            & $8\times 10^{-5}$ & $6\times 10^{-4}$ \\ 
            $\frac{\partial \omega^\mathrm{pole}_1}{\partial p_2}$ & $-4.782 \times 10^{11}$ & $4.9 \times 10^{10}$ 
            & $7\times 10^{-5}$ & $2\times 10^{-3}$ \\ 
            $\omega^\mathrm{zero}_1$ & $1.357\times 10^{15}$ & $1.066\times 10^{13}$ 
            & $< 10^{-8}$ & $3\times 10^{-8}$ \\ 
            $\frac{\partial \omega^\mathrm{zero}_1}{\partial p_1}$ & $-1.26 \times 10^{12}$ & $-2.8 \times 10^{11}$ 
            & $7\times 10^{-4}$ & $3\times 10^{-3}$ \\ 
            $\frac{\partial \omega^\mathrm{zero}_1}{\partial p_2}$ & $ -4.7 \times 10^{11}$ & $ 8.8 \times 10^{10}$ 
            & $2\times 10^{-3}$ & $2\times 10^{-3}$ \\ 
		\mybottomrule
	\end{tabularx}
	\label{tab:table1}
\end{table}

\begin{figure}[]
\includegraphics[width=0.49\textwidth]{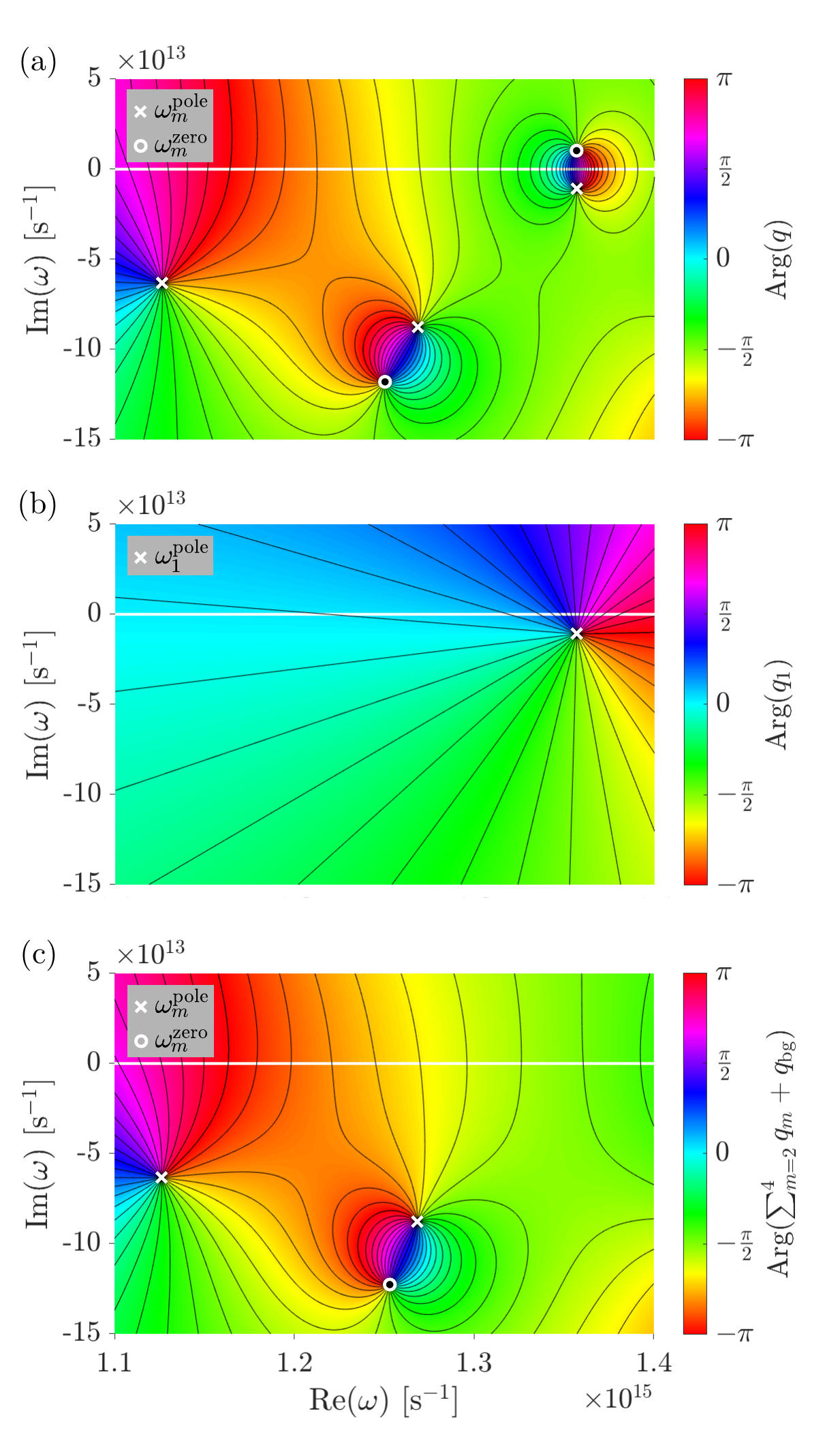}
\caption{\label{fig03}
Phase distribution of the electric field reflected from an illuminated metasurface
and corresponding modal contributions.
(a) Phase $\mathrm{Arg}(q(\omega))$ of the Fourier coefficient $q(\omega)$
in a region enclosed by the contour $C$ shown in Fig.~\ref{fig02}(b).
This phase distribution is obtained by a modal expansion where
$q(\omega)$ is computed only at the integration points $\hat{\omega}_k$ on the contour $C$.
As expected, the locations of the poles $\omega^{\mathrm{pole}}_m$ and zeros
$\omega^{\mathrm{zero}}_m$ from Fig.~\ref{fig02}(b), 
which are marked with white crosses and circles, respectively,
coincide with the poles and zeros of the modal expansion $\mathrm{Arg}(q(\omega))$.
(b) Modal contribution $\mathrm{Arg}(q_1(\omega))$ to the phase.
(c) Sum over remaining phase contributions,
$\mathrm{Arg}(\sum_{m=2}^4 q_m(\omega) + q_\mathrm{bg}(\omega)$).}
\end{figure}

\section{Poles and reflection zeros of a metasurface} We apply this
approach to determine the poles $\omega^{\mathrm{pole}}_{m}$ and reflection zeros
$\omega^{\mathrm{zero}}_{m}$ of the metasurface sketched in Fig.~\ref{fig01}(a).
Figure~\ref{fig02}(a) shows 
the geometry of the nanostructures forming the metasurface, 
including the parameters chosen for the numerical simulation. 
The metasurface is illuminated by a plane wave at normal incidence from above. 
For the investigation of the reflected electric field, we consider the Fourier transform
of $\mathbf{E}(\mathbf{r},\omega_0)$~\cite{Novotny_Hecht_2012}. Due to sub-wavelength periodicity,
the resulting upward propagating Fourier spectrum consists of only one term,
the zero-order diffraction coefficient $Q(\omega_0)$.
Solving the generalized eigenproblem given by Eq.~\eqref{eq:Hankel} with 
the analytical continuation $q(\omega)$ of $Q(\omega_0)$
gives the poles $\omega^{\mathrm{pole}}_m$ and the reflection
zeros $\omega^{\mathrm{zero}}_m$ of the illuminated metasurface.
We emphasize that Eq.~\eqref{eq:Hankel} provides an expression  
of both poles and reflection zeros and that
the numerical implementation does not pose any difficulties.  
Figure~\ref{fig02}(b) shows the integration contour $C$ and the computed poles and zeros.

The contour-integral-based elements of the Hankel matrices in Eq.~\eqref{eq:Hankel}
allow to apply the approach of direct differentiation~\cite{Binkowski_CommunPhys_2022}. 
When the Fourier coefficients $q(\hat{\omega}_k)$ are calculated
at the integration points $\hat{\omega}_k$ on the contour $C$,
also their sensitivities $\partial q/ \partial p$ with respect to geometry, material, or source 
parameters $p$ can be evaluated without significant additional computational effort. 
The sensitivities of the zeros can be extracted in the same way 
as the sensitivities of the poles can be extracted~\cite{Binkowski_CommunPhys_2022}.
Figure~\ref{fig02}(c) sketches the sensitivities $\partial \omega^{\mathrm{pole}}_1/ \partial p_k$
and $\partial \omega^{\mathrm{zero}}_1/ \partial p_k$ with respect
to the upper radius $p_1$ and the height $p_2$ of the silicon cones of the metasurface. 
With $64$ integration points, it is possible to compute poles,
zeros, and their sensitivities with high accuracies, see Table~\ref{tab:table1}.
Numerical convergence with respect to the number of integration points can be observed,
see Sec.~S1 in the Supplemental Material~\cite{Supplement_Poles_Zeros}.
To demonstrate the general applicability of the presented approach, we investigate another
electromagnetic response function, the transmission coefficients
of a photonic structure supporting a bound state in the continuum,
in Sec.~S2 in the Supplemental Material~\cite{Supplement_Poles_Zeros}.

\section{Modal expansion in the complex frequency plane} The residues 
 \begin{align}
    a_m = \frac{1}{2 \pi i} \oint_{C_m} q(\omega)d\omega, \nonumber
 \end{align}
where $C_m$ are contours enclosing the single eigenvalues $\omega_m$ from Eq.~\eqref{eq:Hankel}, 
can be used as a selection criterion for meaningful eigenvalues $\omega_m$.
Eigenvalues with large $a_m$ are prioritized, while $\omega_m$ with small $a_m$ are likely
to be unphysical eigenvalues because either $M$ is chosen larger than the actual number of eigenvalues
within the contour or they are not significant with respect to the quantity of interest.
Correspondingly, the choice of a specific source in Eq.~\eqref{eq:Maxwell} allows to regard only a subset of 
eigenvalues of the considered physical system~\cite{Betz_2021,Binkowski_SourceCode_Poles_Zeros}.
Note that, for simple eigenvalues, the residues are 
directly available, given by $\mathrm{diag}(a_1,\dots,a_M) = X^T H X$,
where $X$ is suitably scaled~\cite{Binkowski_SourceCode_Poles_Zeros}.

Moreover, with the poles $\omega^{\mathrm{pole}}_m$
and the corresponding residues $a_m$, the residue-based modal expansion
of the Fourier coefficient,
\begin{align}
\begin{split}
      & q(\omega) = \sum_{m=1}^{M} q_m(\omega) + q_\mathrm{bg}(\omega), \\
           q_m(\omega) = & \frac{-a_m}{ \omega^{\mathrm{pole}}_m - \omega},\,\,\,
           q_\mathrm{bg}(\omega) = \frac{1}{2 \pi i} \oint_C \frac{q(\xi)}{\xi - \omega}d\xi, \label{eq:modal_exp}
     \end{split}  
\end{align}
can be performed, where 
$q_m(\omega)$ are Riesz-projection-based modal contributions and $q_\mathrm{bg}(\omega)$ is the 
background contribution~\cite{Zschiedrich_PRA_2018}.

Figure~\ref{fig03}(a) shows the phase distribution $\mathrm{Arg}(q(\omega))$ of the electric field
reflected from the metasurface shown in Fig.~\ref{fig02}(a).
This is obtained by evaluating the modal expansion given by Eq.~\eqref{eq:modal_exp}
for the contour $C$ shown in Fig.~\ref{fig02}(b).
A phase retardation of $2\pi$ for a real frequency scan,
which is often required for the design of metasurfaces, 
is obtained when a pair of pole $\omega^\mathrm{pole}_m$ and zero $\omega^\mathrm{zero}_m$ 
is separated by the real axis~\cite{Colom_2023}.
Figure~\ref{fig03}(b) shows $\mathrm{Arg}(q_1(\omega))$ corresponding to the pole $\omega^\mathrm{pole}_1$ and 
Fig.~\ref{fig03}(c) shows $\mathrm{Arg}(\sum_{m=2}^{4} q_m(\omega) + q_\mathrm{bg}(\omega))$.
In particular, it can be observed that the zero $\omega^\mathrm{zero}_1$ does not appear for the 
modal contribution $q_1(\omega)$, but it emerges due to interference
with the other contributions, i.e., when 
$\sum_{m=2}^{4} q_m(\omega) + q_\mathrm{bg}(\omega)$
is added to $q_1(\omega)$.

\section{Conclusion} We presented a theoretical formulation to determine the locations of complex-valued
singularities, including poles and zeros, in non-Hermitian systems. 
The zeros can be determined by contour integration,
in the same way as the poles corresponding to resonances can be computed. 
We also presented residue-based modal expansions in the complex frequency plane of the phase
of the field reflected from a photonic metasurface, where the total expansion validated the 
computed reflection zeros.
The different modal contributions give insight into the emergence of the reflection zeros 
by interference of various expansion terms. 
Furthermore, computation of partial derivatives of the reflection zeros was demonstrated. 
The approach can easily be transferred to other physical systems 
supporting resonances, e.g., to quantum mechanics and acoustics.

The theory essentially relies on detecting singularities of meromorphic functions
in the complex plane.
Therefore, it can be easily applied to compute other 
response functions, e.g., 
$S$-matrix and transmission coefficients, coefficients of the Jones matrix, scattering cross sections
of isolated particles, or maximal chiral response of nanoassemblies. 
The real frequency response of metasurfaces can in many cases be
significantly impacted by reflection and transmission zeros, 
since these typically lie close to the real axis or can even cross
the real axis with slight parameter variations,
see also Sec.~S3 in the Supplemental Material~\cite{Supplement_Poles_Zeros}.
Therefore, a precise quantification of the sensitivities of 
reflection and transmission zeros or also of other physical quantities 
is essential for gradient-based optimization of 
photonic metasurfaces or other non-Hermitian systems.
We expect that the presented theory will enable new computer-aided design approaches.

\section*{Data and code availability}
Supplementary data tables and source code for the numerical experiments
for this work can be found in the open access data publication~\cite{Binkowski_SourceCode_Poles_Zeros}.
\vspace{-0.1cm}

\section*{Acknowledgements}
We acknowledge funding
by the Deutsche Forschungsgemeinschaft (DFG, German Research Foundation) 
under Germany's Excellence Strategy - The Berlin Mathematics Research
Center MATH+ (EXC-2046/1, project ID: 390685689),
by the German Federal Ministry of Education and Research
(BMBF Forschungscampus MODAL, project 05M20ZBM),
 and by the European Innovation Council (EIC) project TwistedNano
(grant agreement number Pathfinder Open 2021-101046424).

\clearpage
\newpage



\section*{Supplementary material}

# Poles and zeros in non-Hermitian systems: Application to photonics
## Supplemental Material

Felix Binkowski,[1] Fridtjof Betz,[1] Rémi Colom,[2] Patrice Genevet,[2,3] and Sven Burger[1,4]

[1]*Zuse Institute Berlin, 14195 Berlin, Germany*
[2]*Université Côte d'Azur, CNRS, CRHEA, 06560 Valbonne, France*
[3]*Physics Department, Colorado School of Mines, Golden, Colorado 80401, USA*
[4]*JCMwave GmbH, 14050 Berlin, Germany*

The main text introduces a framework for the investigation of poles and zeros in non-Hermitian systems, with application to photonics. In this supplemental material, additional technical information is detailed. Numerical results from the main text are validated by comparison to alternative methods, and by a numerical convergence study. An additional application example demonstrates the applicability of the framework to a different photonic response function and to a different problem class. To support a possible application perspective, various design approaches for photonic structures are reviewed. These approaches are based on the locations of poles and zeros of response functions in the complex frequency plane.

## S1. NUMERICAL VALIDATION OF THE RESULTS

The focus of the study in the main text is on the scattered electric field reflected from the metasurface sketched in Fig. 2(a), which is illuminated by a plane wave at normal incidence from above. To compute the contour integrals for the elements of the Hankel matrices of the generalized eigenproblem given by Eq. (2), Maxwell's equation given by Eq. (1) has to be solved for complex-valued frequencies on the integration contour. Maxwell's equation is spatially discretized by the finite element method (FEM). The FEM solver JCMsuite is used to solve the scattering problems [S1], where outgoing radiation conditions are realized by perfectly matched layers. To achieve a sufficient accuracy of the FEM discretization, high-order polynomial ansatz functions and mesh refinements are used. The implementation of the numerical simulation of the specific physical problem and the implementation of the contour-integral-based approach are provided in a data publication [S2].

### A. Comparison with the Arnoldi method

Figure S1(a) shows the complex frequency plane with the integration contour $C$ and with the resulting poles $\omega_m^{\mathrm{pole}}$ and reflection zeros $\omega_m^{\mathrm{zero}}$ from the main text. Furthermore, we compute poles using the Arnoldi method [S3] as implemented in JCMsuite. The Arnoldi method is not based on contour integration and therefore provides independent reference results. The corresponding results $\omega_{\mathrm{ref},m}^{\mathrm{pole}}$ are also shown in Fig. S1(a). Here, we only show the results of the Arnoldi method that lie within the contour. It can be seen that the contour-integral-based approach yields less poles within the contour than the Arnoldi method. This is due to the fact that the contour integration takes into account a certain physical quantity, namely the zero-order diffraction coefficient of the Fourier spectrum propagating upwards. The poles which have much smaller residues than the other poles can be filtered out automatically [S2, S4]. Thus, only poles that contribute to the physical quantity under investigation are obtained, i.e., only relevant poles. The complex frequencies of the relevant poles computed with the contour-integral-based approach agree with four of the results of the Arnoldi method, with a maximum relative error of $10^{-6}$. This validates the computation of poles using the contour-integral-based framework.

### B. Convergence with respect to the number of integration points

As solving the scattering problems at the integration points on the contour takes the major part of the computation time, we show, in Fig. S1(b-g), the relative errors of the poles, reflection zeros, and their sensitivities of the metasurface investigated in the main text with respect to the number of integration points. This provides additional information to the values shown in Table 1 in the main text. It can be observed that, for the poles and zeros, already 32 integration points are sufficient to achieve a relative error smaller than $10^{-6}$. Note that the accuracy with respect to the integration points is limited by the accuracy of the FEM discretization. The numerical convergence properties of the results validate the implementation of the contour-integral-based approach.

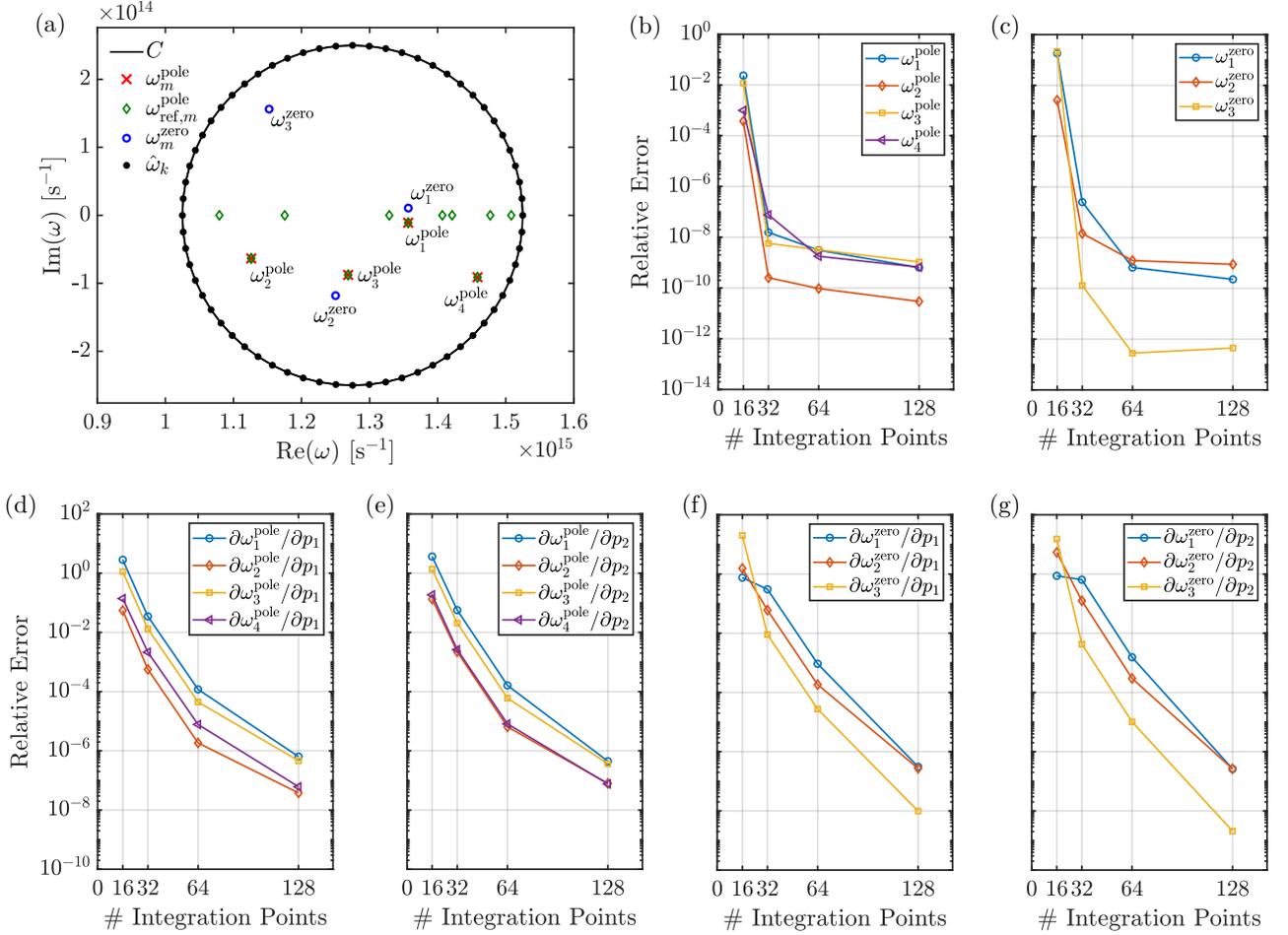

FIG. S1. Convergence study of the poles, reflection zeros, and the corresponding sensitivities of the metasurface from the main text. (a) Integration contour $C$ from the main text with integration points $\hat{\omega}_k$ and the resulting poles $\omega_m^{\text{pole}}$ and reflection zeros $\omega_m^{\text{zero}}$. The Arnoldi method yields the poles $\omega_{\text{ref},m}^{\text{pole}}$. (b) Relative errors $|\omega_m^{\text{pole}} - \omega_{\text{end},m}^{\text{pole}}|/|\omega_{\text{end},m}^{\text{pole}}|)$ of the poles, where $\omega_{\text{end},m}^{\text{pole}}$ is computed using 256 integration points. (c) Relative errors of the reflection zeros. (d-g) Relative errors of the sensitivities of the poles and zeros. The parameters $p_1$ and $p_2$ correspond to the upper radius and to the height of the cones of the metasurface.

## C. Modal expansion of the reflection

For the numerical validation of the modal expansion given by Eq. (3), we consider the reflection $R(\omega_0) = |q(\omega_0)|^2$ for real-valued frequencies $\omega_0 \in \mathbb{R}$, where $q(\omega_0)$ is the Fourier coefficient from the main text. This leads to the modal expansion $R(\omega_0) = \sum_{m=1}^{M} R_m(\omega_0) + R_{\text{bg}}(\omega_0)$, where the modal contributions $R_m(\omega_0)$ and the background contribution $R_{\text{bg}}(\omega_0)$ are given in a similar way as in Eq. (3). Figure S2 shows the expansion, the single modal contributions for the poles $\omega_1^{\text{pole}}, \ldots, \omega_4^{\text{pole}}$, and the background contribution for the reflection $R(\omega_0)$. We compare the modal expansion with the quasiexact solution, where scattering problems for the corresponding real-valued frequencies are solved. The modal expansion agrees with the quasiexact solution up to a maximum relative error of $10^{-3}$. This agreement validates the contour-integral-based modal expansion from the main text.

## S2. POLE AND ZERO OF A BOUND STATE IN THE CONTINUUM

To demonstrate the applicability of the framework presented in the main text, we study another photonic example from the literature [S5], where the interplay between poles and transmission zeros of a metasurface is essential for understanding the underlying physics. In this example, a metasurface with a broken in-plane inversion symmetry of

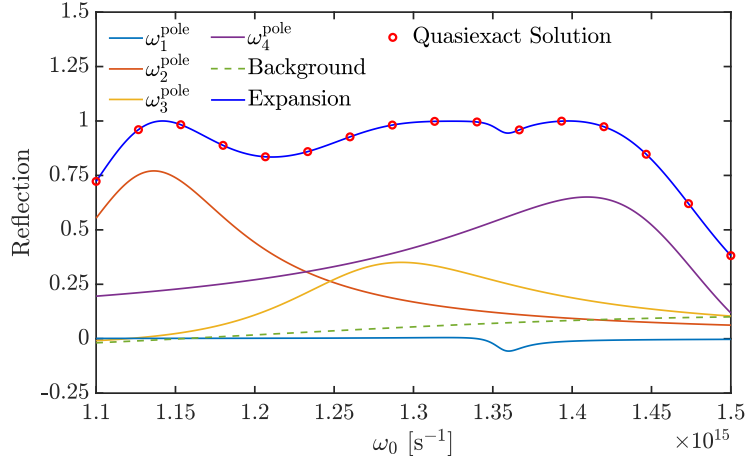

FIG. S2. Modal expansion of the reflection $R(\omega_0) = |q(\omega_0)|^2$ of the metasurface studied in the main text, where the contour $C$ from the main text with 64 integration points is used. The single modal contributions for the poles $\omega_1^{\text{pole}}, \dots, \omega_4^{\text{pole}}$, the background contribution, the total modal expansion, and the quasiexact solution are shown. The expansion coincides with the quasiexact solution, which is given by scattering solutions for real-valued frequencies.

the meta-atoms is considered. This is realized by a square array of silicon-bar pairs tilted by the angle $\Theta$. Changing the asymmetry parameter $\Theta$ in a lossless and infinite system such that the bars are parallel to each other leads to a symmetry-protected bound state in the continuum (BIC).

## A. Pole and transmission zero merge at the real axis

In Fig. S3(a), we recompute the transmission spectra from Ref. [S5] for different values of $\Theta$. The spectra are computed with a modal expansion based on the presented framework from the main text, and with quasiexact solutions based on scattering simulations for real-valued frequencies. The quasiexact solutions have been added as a reference to demonstrate the accuracy. A vanishing transmission peak can be observed when $\Theta$ is varied from 80° to 90°. For $\Theta = 90°$, the metasurface supports a BIC. Figure S3(b) sketches the unit cell of the metasurface with the inclination angle $\Theta$.

Figure S3(c) shows the trajectories of the poles and transmission zeros of the underlying system when the BIC emerges. The poles and transmission zeros corresponding to the BIC coincide on the real axis when the symmetry of the meta-atoms is established. Since we have a system without absorption, the transmission zeros for all $\Theta$ are on the real axis. However, the poles are complex-valued since open boundaries are present. We also compute the sensitivities of the poles and transmission zeros with respect to $\Theta$. The directions of the sensitivities agree with the motion of the poles and zeros when $\Theta$ changes. The poles, zeros, and their sensitivities are computed using the contour-integral-based framework from the main text, based on the contour $C$ shown in Fig. S3(d) with 16 integration points. The numeric values of poles and zeros shown in Fig. S3(c) are provided in Table S1. The numerical implementation of the metasurface and the parameters for the contour integration can be found in Ref. [S2].

## B. Accuracy of the contour integration

In Table S1, poles and transmission zeros computed with the contour $C$ shown in Fig. S3(d) are presented. For inclination angles $\Theta < 89$, only 16 integration points suffice to push the relative error of the numerical integration below $10^{-5}$. The reference solution is computed using 64 integration points. It can been observed that, for increasing $\Theta$, the relative errors increase. This can be explained by the weakening coupling between poles or zeros and the source term for the contour integration. As the meta-atoms approach the symmetric case, the system approaches more and more the BIC, which is more difficult to excite due to the small imaginary part of the underlying pole.

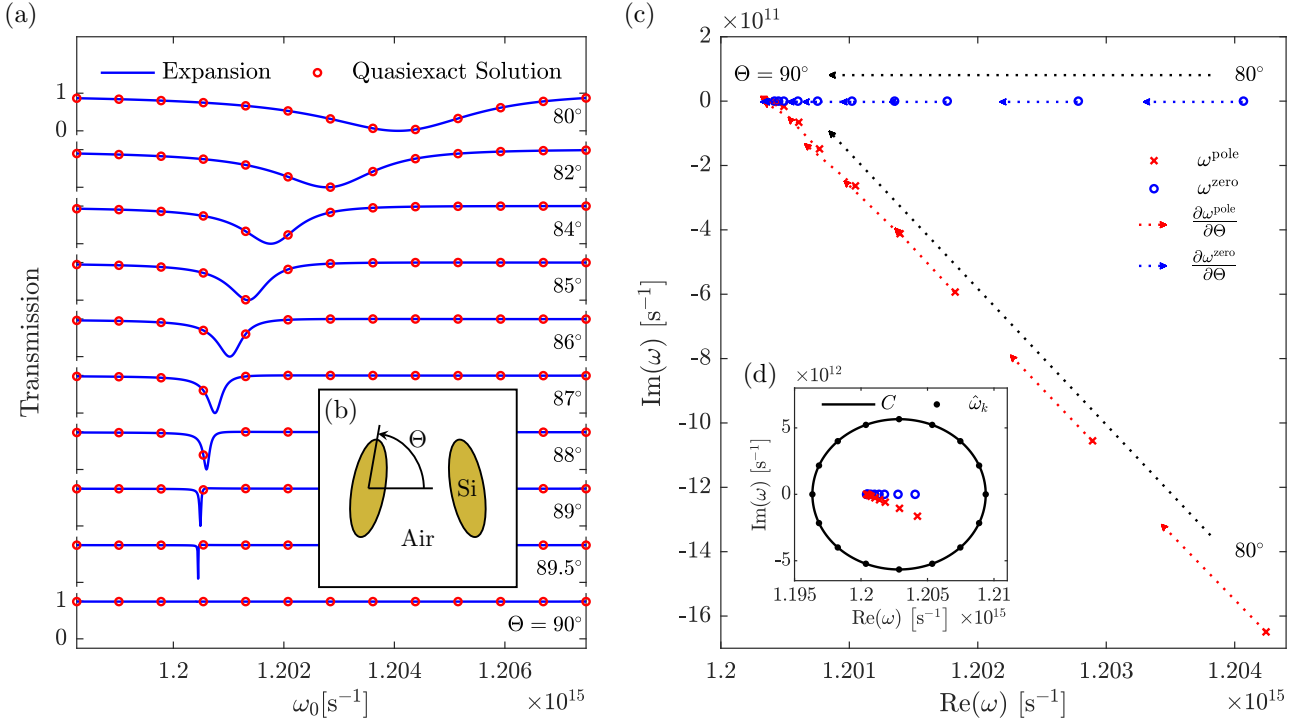

FIG. S3. Metasurface supporting a BIC. (a) Transmission spectra of the metasurface for different inclination angles $\Theta$, computed using modal expansions based on the framework presented in the main text. The expansion can be compared with quasiexact solutions. (b) Schematic of the square unit cell. (c) Complex frequency plane with poles, transmission zeros, and their sensitivities for various $\Theta$. (d) The contour used to compute the results in (a) and (c), with 16 integration points $\hat{\omega}_k$. For comparison, the poles and zeros computed for also are also shown.

## S3.   USE OF POLES AND ZEROS OF RESPONSE FUNCTIONS TO DESIGN PHOTONIC STRUCTURES WITH A DESIRED OPTICAL RESPONSE

The main text is concerned with the development of a framework to find the poles and zeros of response functions, e.g., $S$-matrix, reflection, transmission, or Jones matrix coefficients, of photonic structures. In this section, to support the results from the main text, we review situations where the study of relative locations of poles and zeros in the complex frequency plane helps designing photonic structures, in particular metasurfaces. In the first subsection, we consider design approaches necessitating the study of poles alone. In the second subsection, we discuss design approaches based on the zeros of response functions alone. Eventually, in the third subsection, we consider situations where the design requires the knowledge of the relative locations of poles and zeros. This demonstrates the applicability of the framework developed in the main text for the design of photonic structures in general and metasurfaces in particular.

### A.   Design based on the locations of poles

Here, we only discuss the design of passive photonic systems, i.e., systems without optical gain. As a result, for an $e^{-i\omega t}$ time dependency, the poles of the response functions are restricted to the lower part of the complex frequency plane including the real axis. Indeed, BICs correspond to conditions for which poles and zeros merge on the real axis [S6, S7] as illustrated in the example of Sec. S2. Numerous applications depend on the quality ($Q$) factor of resonant photonic structures. This is the case, e.g., for the control of the emission of quantum emitters or for bio-sensing methods based on a resonance frequency shift. In this context, it is very useful to develop tools allowing to systematically determine the $Q$-factor of the resonances of photonic structures and the dependency on their optical and geometrical parameters. Exceptional points, which correspond to resonance degeneracies specific for non-Hermitian systems, are associated with second order poles of the response functions [S8]. Designing structures

TABLE S1. Poles and transmission zeros for various inclination angles $\Theta$. The results are computed using 16 integration points on the contour $C$ shown in Fig. S3(d). If the relative error is larger than $10^{-6}$, the decimal place with uncertainty is marked with a bracket. As in the main text, the relative error is defined by $\mathrm{err}(\mathrm{Re}(\omega)) = |(\mathrm{Re}(\omega) - \mathrm{Re}(\omega_{\mathrm{ref}}))/\mathrm{Re}(\omega_{\mathrm{ref}})|$, where $\omega_{\mathrm{ref}}$ is the reference solution computed with 64 integration points.

| $\Theta$ | $\mathrm{Re}(\omega^{\mathrm{pole}})$ | $\mathrm{Im}(\omega^{\mathrm{pole}})$ | $\mathrm{Re}(\omega^{\mathrm{zero}})$ | $\mathrm{Im}(\omega^{\mathrm{zero}})$ |
|---|---|---|---|---|
| 90 | $1.20042 \times 10^{15}$ | $-2.(9) \times 10^5$ | $1.20042 \times 10^{15}$ | $2.(8) \times 10^7$ |
| 89.5 | $1.20045 \times 10^{15}$ | $-4.3(5) \times 10^9$ | $1.20045 \times 10^{15}$ | $2.64712 \times 10^7$ |
| 89 | $1.20049 \times 10^{15}$ | $-1.638(5) \times 10^{10}$ | $1.20049 \times 10^{15}$ | $2.48022 \times 10^7$ |
| 88 | $1.20061 \times 10^{15}$ | $-6.57025 \times 10^{10}$ | $1.20060 \times 10^{15}$ | $2.49674 \times 10^7$ |
| 87 | $1.20077 \times 10^{15}$ | $-1.48385 \times 10^{11}$ | $1.20075 \times 10^{15}$ | $2.88135 \times 10^7$ |
| 86 | $1.20105 \times 10^{15}$ | $-2.63303 \times 10^{11}$ | $1.20102 \times 10^{15}$ | $2.81287 \times 10^7$ |
| 85 | $1.20140 \times 10^{15}$ | $-4.12009 \times 10^{11}$ | $1.20135 \times 10^{15}$ | $2.81741 \times 10^7$ |
| 84 | $1.20182 \times 10^{15}$ | $-5.93235 \times 10^{11}$ | $1.20176 \times 10^{15}$ | $2.78677 \times 10^7$ |
| 82 | $1.20289 \times 10^{15}$ | $-1.05550 \times 10^{12}$ | $1.20278 \times 10^{15}$ | $2.76697 \times 10^7$ |
| 80 | $1.20424 \times 10^{15}$ | $-1.64924 \times 10^{12}$ | $1.20407 \times 10^{15}$ | $2.73544 \times 10^7$ |

supporting second order poles necessitates to fine tune their geometry. In this context, the finding of the poles and the determination of their sensitivities with respect to some parameters is useful.

### B. Design based on the locations of zeros

It is also possible to design photonic structures with a desired optical response by considering the zeros of response functions. The zeros may be generally complex-valued, but interesting behaviors occur in structures for which these zeros are real-valued. It has long been known that the zeros of the $S$-matrix coefficients of lossless photonic structures are complex conjugates of their poles due to time-reversal symmetry and are therefore complex-valued. Forcing these zeros to be real-valued by adding the right amount of absorptive loss in the structures leads to total absorption of light. If the structures are on an absorptive substrate, then a plane wave incident from above can be totally absorbed when a zero of the $S$-matrix is real-valued [S9]. Perfect coherent absorption may also occur by shining coherent beams from both sides of a photonic structure, e.g., an absorptive slab [S10]. Reflectionless scattering states occur when the zeros of the reflection coefficients are real-valued [S11]. It has been shown that reflectionless states occurring for real-valued frequencies are related to the PT-symmetry of the considered photonic structure [S11, S12]. Transmissionless states, occurring when transmission zeros are real-valued, are also very important for the design of photonic structures [S13, S14]. The behavior of the phase around zeros of Jones matrix coefficients have been used to realize phase-gradient metasurfaces [S15, S16]. Finally, structures with second order zeros, which are another kind of exceptional points, possess unique behaviors opening new approaches for the realization of photonic devices. This has been already illustrated in the literature for the zeros of the $S$-matrix coefficients [S17, S18], the reflection zeros [S19], and transmission zeros [S14].

### C. Design based on the relative locations of poles and zeros

Finally, there are also situations where the design of photonic structures necessitates considering a pair of pole and zero. BICs are a striking example as they arise when a pole and a zero merge on the real axis as demonstrated in Sec. S2. The realization of metalenses or metadeflectors relies on the use of metasurfaces covering the entire range of induced phase shifts from 0 to $2\pi$ to generate a spatial phase gradient. Recently, it has been shown that a sufficient condition for a resonant metasurface to cover the full $2\pi$ interval is to have a pole-zero pair separated by the real axis. Poles are usually confined to the lower part of the complex plane for passive metasurfaces as explained above. This implies that a zero has to be in the upper part of the complex plane [S14]. This condition is general as it applies to the reflection and to the transmission, as well as to the $S$-matrix coefficients. Moreover, a desired behavior of the

amplitude of the response functions may also depend on the location of both poles and zeros. In particular, it has been shown in the Refs. [S12, S20] that a large and constant amplitude may be obtained when poles and zeros are complex conjugates of each other. The example shown in the main text is chosen because it possesses a pole-zero pair which is complex-conjugated around $1.35 \times 10^{15} \text{s}^{-1}$ in Fig. 2(b).



\end{document}